\documentstyle[12pt]{article}

\catcode`\@=11
\long\def\@makefntext#1{ 
\protect\noindent \hbox to 3.2pt {\hskip-.9pt
$^{{\ninerm\@thefnmark}}$\hfil}#1\hfill} 

\def\thefootnote{\fnsymbol{footnote}}
 \def\@makefnmark{\hbox to 0pt{$^{\@thefnmark}$\hss}}  

\def\ps@myheadings{\let\@mkboth\@gobbletwo
\def\@oddhead{\hbox{} 
\rightmark\hfil\ninerm\thepage}
\def\@oddfoot{}\def\@evenhead{\ninerm\thepage\hfil 
\leftmark\hbox{}}\def\@evenfoot{}
\def\sectionmark##1{}\def\subsectionmark##1{}}

\begin{document}

\newcommand{\symbolfootnote}{\renewcommand{\thefootnote}
        {\fnsymbol{footnote}}}
\renewcommand{\thefootnote}{\fnsymbol{footnote}}
\newcommand{\alphfootnote}
        {\setcounter{footnote}{0}
         \renewcommand{\thefootnote}{\sevenrm\alph{footnote}}}

\newcounter{sectionc}\newcounter{subsectionc}\newcounter{subsubsectionc}
\renewcommand{\section}[1] {\vspace{0.6cm}\addtocounter{sectionc}{1}
\setcounter{subsectionc}{0}\setcounter{subsubsectionc}{0}\noindent
        {\bf\thesectionc. #1}\par\vspace{0.4cm}}
\renewcommand{\subsection}[1] {\vspace{0.6cm}\addtocounter{subsectionc}{1}
        \setcounter{subsubsectionc}{0}\noindent
        {\it\thesectionc.\thesubsectionc. #1}\par\vspace{0.4cm}}
\renewcommand{\subsubsection}[1]
{\vspace{0.6cm}\addtocounter{subsubsectionc}{1}
        \noindent {\rm\thesectionc.\thesubsectionc.\thesubsubsectionc.
        #1}\par\vspace{0.4cm}}
\newcommand{\nonumsection}[1] {\vspace{0.6cm}\noindent{\bf #1}
        \par\vspace{0.4cm}}

\newcounter{appendixc}
\newcounter{subappendixc}[appendixc]
\newcounter{subsubappendixc}[subappendixc]
\renewcommand{\thesubappendixc}{\Alph{appendixc}.\arabic{subappendixc}}
\renewcommand{\thesubsubappendixc}
        {\Alph{appendixc}.\arabic{subappendixc}.\arabic{subsubappendixc}}

\renewcommand{\appendix}[1] {\vspace{0.6cm}
        \refstepcounter{appendixc}
        \setcounter{figure}{0}
        \setcounter{table}{0}
        \setcounter{equation}{0}
        \renewcommand{\thefigure}{\Alph{appendixc}.\arabic{figure}}
        \renewcommand{\thetable}{\Alph{appendixc}.\arabic{table}}
        \renewcommand{\theappendixc}{\Alph{appendixc}}
        \renewcommand{\theequation}{\Alph{appendixc}.\arabic{equation}}
        \noindent{\bf Appendix \theappendixc #1}\par\vspace{0.4cm}}
\newcommand{\subappendix}[1] {\vspace{0.6cm}
        \refstepcounter{subappendixc}
        \noindent{\bf Appendix \thesubappendixc. #1}\par\vspace{0.4cm}}
\newcommand{\subsubappendix}[1] {\vspace{0.6cm}
        \refstepcounter{subsubappendixc}
        \noindent{\it Appendix \thesubsubappendixc. #1}
        \par\vspace{0.4cm}}

\def\abstracts#1{{
        \centering{\begin{minipage}{30pc}\tenrm\baselineskip=12pt\noindent
        \centerline{\tenrm ABSTRACT}\vspace{0.3cm}
        \parindent=0pt #1
        \end{minipage} }\par}}

\newcommand{\bibit}{\it}
\newcommand{\bibbf}{\bf}
\renewenvironment{thebibliography}[1]
        {\begin{list}{\arabic{enumi}.}
        {\usecounter{enumi}\setlength{\parsep}{0pt}
\setlength{\leftmargin 1.25cm}{\rightmargin 0pt}
         \setlength{\itemsep}{0pt} \settowidth
        {\labelwidth}{#1.}\sloppy}}{\end{list}}

\topsep=0in\parsep=0in\itemsep=0in
\parindent=1.5pc

\newcounter{itemlistc}
\newcounter{romanlistc}
\newcounter{alphlistc}
\newcounter{arabiclistc}
\newenvironment{itemlist}
        {\setcounter{itemlistc}{0}
         \begin{list}{$\bullet$}
        {\usecounter{itemlistc}
         \setlength{\parsep}{0pt}
         \setlength{\itemsep}{0pt}}}{\end{list}}

\newenvironment{romanlist}
        {\setcounter{romanlistc}{0}
         \begin{list}{$($\roman{romanlistc}$)$}
        {\usecounter{romanlistc}
         \setlength{\parsep}{0pt}
         \setlength{\itemsep}{0pt}}}{\end{list}}

\newenvironment{alphlist}
        {\setcounter{alphlistc}{0}
         \begin{list}{$($\alph{alphlistc}$)$}
        {\usecounter{alphlistc}
         \setlength{\parsep}{0pt}
         \setlength{\itemsep}{0pt}}}{\end{list}}

\newenvironment{arabiclist}
        {\setcounter{arabiclistc}{0}
         \begin{list}{\arabic{arabiclistc}}
        {\usecounter{arabiclistc}
         \setlength{\parsep}{0pt}
         \setlength{\itemsep}{0pt}}}{\end{list}}

\newcommand{\fcaption}[1]{
        \refstepcounter{figure}
        \setbox\@tempboxa = \hbox{\tenrm Fig.~\thefigure. #1}
        \ifdim \wd\@tempboxa > 6in
           {\begin{center}
        \parbox{6in}{\tenrm\baselineskip=12pt Fig.~\thefigure. #1 }
            \end{center}}
        \else
             {\begin{center}
             {\tenrm Fig.~\thefigure. #1}
              \end{center}}
        \fi}

\newcommand{\tcaption}[1]{
        \refstepcounter{table}
        \setbox\@tempboxa = \hbox{\tenrm Table~\thetable. #1}
        \ifdim \wd\@tempboxa > 6in
           {\begin{center}
        \parbox{6in}{\tenrm\baselineskip=12pt Table~\thetable. #1 }
            \end{center}}
        \else
             {\begin{center}
             {\tenrm Table~\thetable. #1}
              \end{center}}
        \fi}

\def\@citex[#1]#2{\if@filesw\immediate\write\@auxout
        {\string\citation{#2}}\fi
\def\@citea{}\@cite{\@for\@citeb:=#2\do
        {\@citea\def\@citea{,}\@ifundefined
        {b@\@citeb}{{\bf ?}\@warning
        {Citation `\@citeb' on page \thepage \space undefined}}
        {\csname b@\@citeb\endcsname}}}{#1}}

\newif\if@cghi
\def\cite{\@cghitrue\@ifnextchar [{\@tempswatrue
        \@citex}{\@tempswafalse\@citex[]}}
\def\citelow{\@cghifalse\@ifnextchar [{\@tempswatrue
        \@citex}{\@tempswafalse\@citex[]}}
\def\@cite#1#2{{$\null^{#1}$\if@tempswa\typeout
        {IJCGA warning: optional citation argument
        ignored: `#2'} \fi}}
\newcommand{\citeup}{\cite}

\def\fnm#1{$^{\mbox{\scriptsize #1}}$}
\def\fnt#1#2{\footnotetext{\kern-.3em
        {$^{\mbox{\sevenrm #1}}$}{#2}}}

\font\twelvebf=cmbx10 scaled\magstep 1
\font\twelverm=cmr10 scaled\magstep 1
\font\twelveit=cmti10 scaled\magstep 1
\font\elevenbfit=cmbxti10 scaled\magstephalf
\font\elevenbf=cmbx10 scaled\magstephalf
\font\elevenrm=cmr10 scaled\magstephalf
\font\elevenit=cmti10 scaled\magstephalf
\font\bfit=cmbxti10
\font\tenbf=cmbx10
\font\tenrm=cmr10
\font\tenit=cmti10
\font\ninebf=cmbx9
\font\ninerm=cmr9
\font\nineit=cmti9
\font\eightbf=cmbx8
\font\eightrm=cmr8
\font\eightit=cmti8

\newcommand{\bea}{\begin{eqnarray}}
\newcommand{\eea}{\end{eqnarray}}
\newcommand{\be}{\begin{equation}}
\newcommand{\ee}{\end{equation}}
\newcommand{\nl}{\nonumber \\}
\newcommand{\Av}{{\bf A}}
\newcommand{\Bv}{{\bf B}}
\newcommand{\Dt}{\tensor{D}}
\newcommand{\Dv}{{\bf D}}
\newcommand{\Dtv}{\tensor{\bf D}}
\newcommand{\Ev}{{\bf E}}
\newcommand{\gammav}{\mbox{\boldmath $\gamma$}}
\newcommand{\kv}{{\bf k}}
\newcommand{\lv}{{\bf l}}
\newcommand{\Pv}{{\bf P}}
\newcommand{\pv}{{\bf p}}
\newcommand{\phat}{\hat{\bf p}}
\newcommand{\pvp}{{\bf p'}}
\newcommand{\pvps}{{\bf p'\,}^2}
\newcommand{\qhat}{{\hat {\bf q}}}
\newcommand{\qv}{{\bf q}}
\newcommand{\rv}{{\bf r}}
\newcommand{\rhat}{{\hat {\bf r}}}
\newcommand{\vv}{{\bf v}}
\newcommand{\xv}{{\bf x}}
\newcommand{\sigmav}{\mbox{\boldmath $\sigma$}}
\newcommand{\epsilonv}{\mbox{\boldmath$\epsilon$}}
\newcommand{\gradv}{\mbox{\boldmath $\nabla$}}
\newcommand{\gradt}{\tensor{\mbox{\boldmath $\nabla$}}}
\newcommand{\Tr}{{\rm Tr\,}}
\newcommand{\bra}[1]{{\langle #1 |}}
\newcommand{\ket}[1]{{| #1 \rangle}}
\newcommand{\chidag}{\chi^\dagger}
\newcommand{\psidag}{\psi^\dagger}
\newcommand{\Psibar}{\overline{\Psi}}
\newcommand{\chic}[1]{\chi_{c#1}}
\newcommand{\singS}{{}^1S_0}
\newcommand{\tripS}{{}^3S_1}
\newcommand{\singP}{{}^1P_1}
\newcommand{\tripP}[1]{{}^3P_{#1}}
\newcommand{\singD}{{}^1D_2}
\newcommand{\tripD}[1]{{}^3D_{#1}}
\newcommand{\Reta}{{\overline {R_{\eta_c}}}}
\newcommand{\Rpsi}{{\overline {R_\psi}}}
\newcommand{\RS}{{\overline {R_S}}}
\newcommand{\ddReta}{{\overline {\nabla^2 R_{\eta_c}}}}
\newcommand{\ddRpsi}{{\overline {\nabla^2 R_\psi}}}
\newcommand{\ddRS}{{\overline {\nabla^2 R_S}}}
\newcommand{\DDReta}{{\overline {D^2 R_{\eta_c}}}}
\newcommand{\DDRpsi}{{\overline {D^2 R_\psi}}}
\newcommand{\DDRS}{{\overline {D^2 R_S}}}
\newcommand{\dRh}{{\overline {R_{h_c}'}}}
\newcommand{\dRchi}[1]{{\overline {R_{\chi_{c#1}}'}}}
\newcommand{\dRP}{{\overline {R_P'}}}
\newcommand{\RoctS}{{\overline {R_{{\underline 8}S}}}}
\newcommand{\qbar}{\overline{q}}
\newcommand{\Qbar}{\overline{Q}}
\newcommand{\QQ}{Q \overline{Q}}
\newcommand{\QQg}{Q \overline{Q} g}
\newcommand{\QQgg}{Q \overline{Q} g g}
\newcommand{\cc}{{\rm c.c.}}
\newcommand{\hc}{{\rm h.c.}}
\newcommand{\K}{{\cal K}}
\newcommand{\M}{{\cal M}}
\newcommand{\EM}{{\rm EM}}
\newcommand{\LH}{{\rm LH}}
\newcommand{\gam}{\gamma \gamma}
\newcommand{\ggg}{3 \gamma}
\newcommand{\lep}{\rm ee}
\newcommand{\IR}{{\rm IR}}
\newcommand{\UV}{{\rm UV}}
\newcommand{\MSbar}{\overline{{\rm MS}}}
\newcommand{\NRQCD}{{\rm NRQCD}}
\newcommand{\half}{\mbox{$\frac{1}{2}$}}
\newcommand{\ihalf}{\mbox{$\frac{i}{2}$}}
\newcommand{\fourth}{\mbox{$\frac{1}{4}$}}


\rightline{\vbox{
\halign{&#\hfil\cr
&NUHEP-TH-94-22 \cr
&September 1994 \cr
&hep-ph/9409286 \cr}}}
\bigskip
\bigskip

\centerline{\tenbf RADIATIVE CORRECTIONS TO QUARKONIUM DECAYS:}
\baselineskip=16pt
\centerline{\tenbf FROM A MODEL TO A RIGOROUS THEORY}
\vspace{0.8cm}
\centerline{\tenrm ERIC BRAATEN\footnotemark[1]}
\footnotetext[1]{invited talk at the Tennessee International Symposium on
	Radiative Corrections, Gatlinburg, Tennessee, June 1994.}
\baselineskip=13pt
\centerline{\tenit Department of Physics and Astronomy, Northwestern
University}
\baselineskip=12pt
\centerline{\tenit Evanston, IL 60208, USA}
\vspace{0.9cm}
\abstracts{
Most previous calculations of the annihilation decay rates of heavy quarkonium
were based on factorization assumptions that were unproven and,
in some cases, even incorrect.  The recent development of a general
factorization formula for heavy quarkonium annihilation rates
has provided a rigorous theoretical foundation for such calculations.
The factorization formula is based on the use of the effective field theory
NRQCD to factor the decay rate into short-distance coefficients that can be
calculated in perturbation theory and long-distance matrix elements
that can be computed using lattice simulations.  This approach allows
annihilation decay rates to be computed entirely from first principles.
}

\vfil
\twelverm   
\baselineskip=14pt

\section{History of Quarkonium Annihilation Calculations}

\vspace*{-0.7cm}
\subsection{Leading-order Calculations (1974-1976)}
\vspace*{-0.35cm}

The discovery of charmonium in November 1974 launched a revolution in
particle physics.  Within days of the announcement of the discovery,
there were papers submitted to Physical Review Letters\cite{aprg}
which interpreted the new resonances as bound states of a charm
quark and antiquark analogous to positronium.  Among other things, the authors
calculated the annihilation decay rates of the S-wave states of charmonium,
which were eventually named $\eta_c$ and $J/\psi$.
They assumed that charmonium decays into light hadrons via the annihilation
of the $c \bar c$ pair into gluons, just as positronium decays by
annihilation into photons.  The decay rates for positronium are
proportional to $|R_S(0)|^2$, where $R_S(0)$ is the radial wavefunction
at the origin.  The decay rates for $\eta_c \to g g$,
$\eta_c \to \gamma \gamma$, $\psi \to g g g$, and $\psi \to e^+ e^-$
can all be obtained to leading order in $\alpha_s$
from the corresponding positronium calculations
by replacing $\alpha$ by $\alpha_s$ and $m_e$ by $M$ (where $M$ is
the charm quark mass) in the coefficient of $|R_S(0)|^2$
and multiplying by an overall color factor.

The annihilation decay rates of P-wave charmonium states were calculated
to leading order in $\alpha_s$ by Barbieri and
collaborators\cite{barbb,barbc} in 1976.  The decay rates were assumed to be
proportional to $|R_P'(0)|^2$, where $R_P'(0)$ is the derivative of the
radial wavefunction at the origin.  This assumption seemed to be valid for
the decay rates of the spin-0 and spin-2 states into light hadrons via
$\chi_{c0} \to g g$ and $\chi_{c2} \to g g$.  It also seemed to be valid
for their  electromagnetic decays $\chi_{c0} \to \gamma \gamma$ and
$\chi_{c2} \to \gamma \gamma$.  However, for decays of the spin-1
states into light hadrons, the partial decay rates for
$h_c \to g g g$ and $\chi_{c1} \to q {\bar q} g$ were
infrared divergent.  These divergences were
interpreted as a signal that the decay rates are sensitive to
the binding energy and therefore to nonperturbative effects of QCD.

In 1976, the ITEP group\cite{itep}
wrote a review of applications of QCD to charmonium,
including the calculations of annihilation decay rates.  While there
was dramatic improvement in experimental measurements of the decay rates
in subsequent years, their theoretical discussion of charmonium decay rates
remained for the most part up-to-date for over 15 years.  Implicit in the
calculations of the decay rates is the assumption that they can be factored
into a nonperturbative part involving the wavefunction and a
perturbative part, which can be calculated
in terms of the running coupling constant $\alpha_s(M)$.

\vspace*{-0.35cm}
\subsection{Next-to-leading-order Calculations (1979-1981)}
\vspace*{-0.35cm}

The first calculation of the radiative corrections for quarkonium decays
was carried out in 1979 by Barbieri {\it et al.}\cite{barba}.
They calculated the next-to-leading order corrections to the
decay rates for  $\eta_c \to \gamma \gamma$ and $\eta_c \to \LH$, where $\LH$
represents light hadrons.  The next-to-leading order radiative correction for
$\psi \to \LH$ was calculated by Mackenzie and Lepage\cite{ml} in 1981.
These calculations were based on the assumption that
the decay rate could be factored into $|R_S(0)|^2$
multiplied by a perturbative coefficient.
There was no general theory guaranteeing such a factored form,
but the results of these calculations provide empirical evidence
that the assumed factorization occurs,
at least to next-to-leading order in $\alpha_s$.

The next-to-leading order corrections to the decay rates of the P-wave states
$\chi_{c0}$ and $\chi_{c2}$ were calculated by Barbieri and
collaborators\cite{barbe} in 1980.  The decay rates were assumed to
factor into  $|R_P'(0)|^2$ multiplied by a perturbative coefficient.
For the electromagnetic
decays $\chi_{c0} \to \gamma \gamma$ and $\chi_{c2} \to \gamma \gamma$,
the next-to-leading order corrections were consistent with the factorization
assumption.  However, for the inclusive decays into light hadrons
$\chi_{c0} \to \LH$ and $\chi_{c2} \to \LH$, the next-to-leading order
correction was infrared divergent.  As in the case of the decays of
$h_c$ and $\chi_{c1}$ at leading order, these divergences were interpreted
as a signal that the decay rates were sensitive to nonperturbative effects.

\vspace*{-0.35cm}
\subsection{The Breakthrough (1992-1994)}
\vspace*{-0.35cm}

In the decade following the calculations of the radiative corrections
for the S-wave and P-wave states, there was very little progress in the theory
of heavy quarkonium annihilation.  The  breakthrough came in 1992 with the
solution of the problem of infrared divergences in the P-wave decay
rates\cite{bbla}.
The breakthrough was stimulated in part by Experiment E760 at Fermilab,
which produced charmonium states via resonant $p \bar p$-annihilation
and measured their masses, widths, and branching fractions
with unprecedented precision\cite{E760}.
The breakthrough was made possible by the development of
nonrelativistic QCD\cite{caswell-lepage}, an effective field theory for
bound states of heavy quarks and antiquarks.
The NRQCD approach has taken us far beyond the solution to the
P-wave problem.  It provides a general factorization formula
for annihilation decay rates of heavy quarkonium,
including not only perturbative corrections
but relativistic corrections as well\cite{bblb}.  It places the entire
subject of heavy quarkonium annihilation on a firm theoretical foundation.

\section{Problems in the Conventional Approach}

\vspace*{-0.7cm}
\subsection{S-wave Decays}
\vspace*{-0.35cm}

Previous calculations of the annihilation decay rates of heavy quarkonium
were based on the assumption that the decay rate factors into a
short-distance part, which can be calculated as a perturbation series in the
running coupling constant $\alpha_s(M)$, and a long-distance factor,
which depends on the nonperturbative dynamics of the bound state.
To illustrate how such calculations are
carried out in practice, we review the calculation of the decay rate of
the $\eta_c$ to next-to-leading order in $\alpha_s$.

The $\eta_c$ is predominantly a $\ket{c \bar c}$ state, and it decays into
light hadrons through the annihilation of the $c \bar c$ pair.
The annihilation rate depends on the wavefunction of the $c \bar c$ pair,
which is essentially nonperturbative.
Instead of trying to compute the decay rate of the bound state directly,
we calculate the decay rate of a simpler system in perturbation
theory.  This system consists of a $c$ and $\bar c$ that are above threshold
and on their mass-shells, with a prescribed nonrelativistic wavefunction.
The wavefunction is color-singlet ($\delta^{ij}/\sqrt{3}$),
spin-singlet ($(\uparrow \downarrow - \downarrow \uparrow)/\sqrt{2}$), and
S-wave ($\psi({\bf r}) = R(r)/\sqrt{4 \pi}$, with $\int r^2 |R(r)|^2 dr = 1$).
At leading order in perturbation theory, the $c \bar c$ annihilates into
2 gluons.  At next-to-leading order in $\alpha_s$, we must add the rates for
$c \bar c \to g g g$ and $c \bar c \to q \bar q g$, as well as the
interference term between the amplitudes of order $\alpha_s$ and $\alpha_s^2$
for $c \bar c \to g g$.  Ultraviolet divergences are removed by
renormalization of the coupling constant.  Taking the nonrelativistic
limit, we find that the decay rate is proportional to the
square of the radial wavefunction at the origin $R(0)$:
\be
\Gamma(c \bar c) \;=\;
{2 \alpha_s(2M)^2 |R(0)|^2 \over 3 M^2}
\left[ 1 \;+\; {\alpha_s \over \pi} \left( {\pi^2 \over 3} \langle 1/v \rangle
	 + {572 - 31 \pi^2 \over 24} \right) \right]
\label{Getapt}
\ee
The term $\langle 1/v \rangle$ is the average value of the inverse of
the velocity when weighted by the wavefunction.  This term, which
comes from Coulomb scattering of the $c$ and $\bar c$, diverges
in the nonrelativistic limit.  Like $R(0)$, it is sensitive to
nonperturbative effects involving distances much longer than $1/M$.
We can factor the decay rate (\ref{Getapt}) into long-distance and
short-distance parts as follows:
\be
\Gamma(c \bar c) \;=\;
|R(0)|^2 \left( 1 \;+\; {\pi \alpha_s \over 3} \langle 1/v \rangle \right)
\;\times\; {2 \alpha_s(2M)^2 \over 3 M^2}
	\left( 1 \;+\;  {572 - 31 \pi^2 \over 24} {\alpha_s \over \pi} \right)
\label{Getafac}
\ee

We now return to the $\eta_c$, which also decays via the annihilation of the
$c \bar c$ pair.  We assume that this decay rate can also be factored into
short-distance and long-distance parts and that the short-distance part
is insensitive to whether the $c$ and  $\bar c$ are slightly above
threshold and on their mass shells, as in the calculation  above,
or slightly below threshold in a bound state.  The decay rate is then given
by an expression analogous to (\ref{Getapt}), with the same short-distance
factor but a different long-distance factor:
\be
\Gamma(\eta_c \to \LH) \;=\;
|R_S(0)|^2
\;\times\; {2 \alpha_s(2M)^2 \over 3 M^2}
	\left( 1 \;+\; {572 - 31 \pi^2 \over 24} {\alpha_s \over \pi} \right)
\label{Getabarb}
\ee
This next-to-leading order result was first obtained by Barbieri
{\it et al.}\cite{barba} in 1979.
The correction $(\pi \alpha_s/3) \langle 1/v \rangle$,
which arises from Coulomb scattering of the $c \bar c$ pair, has been
factored into $R_S(0)$.  This term is universal, appearing also in the
next-to-leading order corrections for $\eta_c \to \gamma \gamma$,
$\psi \to \LH$, and $\psi \to e^+ e^-$.  Thus one can consistently equate
the long-distance factor $|R_S(0)|^2$ in all of the S-wave decay rates.

\vspace*{-0.35cm}
\subsection{P-wave Decays}
\vspace*{-0.35cm}

Given the success of the factorization assumption for S-wave decays,
it is natural to assume that P-wave decays factor in a similar way.
Instead of $|R_S(0)|^2$, the corresponding long-distance factor would
be $|R_P'(0)|^2$, where $R_P'(0)$ is the derivative of the wavefunction
at the origin.  Using this factorization assumption, the decay rates of
the four P-wave states into light hadrons were calculated by Barbieri
and various collaborators\cite{barbb,barbc,barbe,barbd} to order $\alpha_s^3$.
In the case of $\chi_{c0}$ and $\chi_{c2}$, this is next-to-leading order
in $\alpha_s$.  The next-to-leading order corrections include a
Coulomb singularity $(4 \pi \alpha_s/3) \langle 1/v \rangle$
that can be factored into $|R_P'(0)|^2$.  The resulting expressions for
the decay rates have the form
\bea
\Gamma(h_c \to \LH) &=&
{\alpha_s^2 |R_P'(0)|^2 \over M^4}
\left[ \; 0 \;+\; {\alpha_s \over \pi}
	\left( {20 \over 9} \log {M \over \Lambda} + \ldots \right) \right]
\label{Ghcbarb}
\\
\Gamma(\chi_{c0} \to \LH) &=&
{\alpha_s^2 |R_P'(0)|^2 \over M^4}
\left[ \; 6 \;+\; {\alpha_s \over \pi}
	\left( {8 n_f \over 9} \log {M \over \Lambda} + \ldots \right) \right]
\label{Gchi0barb}
\\
\Gamma(\chi_{c1} \to \LH) &=&
{\alpha_s^2 |R_P'(0)|^2 \over M^4}
\left[ \; 0 \;+\; {\alpha_s \over \pi}
	\left( {8 n_f \over 9} \log {M \over \Lambda} + \ldots \right) \right]
\label{Gchi1barb}
\\
\Gamma(\chi_{c2} \to \LH) &=&
{\alpha_s^2 |R_P'(0)|^2 \over M^4}
\left[ \; {8 \over 5} \;+\; {\alpha_s \over \pi}
	\left( {8 n_f \over 9} \log {M \over \Lambda} + \ldots \right) \right]
\label{Gchi2barb}
\eea
where $n_f=3$ is the number of light quark flavors, the $\ldots$ represent
numerical constants, and $\Lambda$ is an infrared cutoff on the energies of
gluons in the final state.  Logarithmic infrared divergences
appear as we take the limit $\Lambda \to 0$.  It is clear from the
expressions in (\ref{Ghcbarb})--(\ref{Gchi2barb}) that the
infrared divergences
can not be factored into $|R_P'(0)|^2$.  In phenomenological applications,
the infrared cutoff $\Lambda$ has been identified with the inverse of the
radius of the bound state, or with its binding energy, or with the inverse
of the confinement radius.  It is evident, however, that the presence of
infrared divergences implies a failure of the factorization assumption
upon which the entire calculation is based.

\vspace*{-0.35cm}
\subsection{Questions}
\vspace*{-0.35cm}

The explicit calculations of quarkonium annihilation rates carried out
from 1974 to 1981 raise a number of important questions that must be
answered by a rigorous theory of annihilation.

1.  {\em Do the annihilation rates factorize?}
Empirically, factorization holds to next-to-leading order in $\alpha_s$ for
S-waves and also for electromagnetic decays of P-waves.  Does it hold to higher
order?  Empirically, factorization fails for the decays of P-waves into
light hadrons at order $\alpha_s^3$.  Why?

2.  {\em Is there a rigorous field-theoretic definition of the long-distance
factors?}
How do you compute $|R_S(0)|^2$ and $|R_P'(0)|^2$ using lattice simulations?
The long-distance factor $|R_S(0)|^2$ was assumed to be the same for
the decays of the $^1S_0$ and $^3S_1$ states.  It was also assumed to be the
same for decays into light hadrons and decays into electromagnetic final
states.  At what point do these assumptions break down?

3.  {\em How do you calculate relativistic corrections?}
Since $v^2 \approx 1/3$ for charmonium, the relativistic corrections
may be more important than the next-to-leading order perturbative correction.
For S-waves, one would expect the first relativistic correction to have
a long-distance factor that involves
$\nabla^2 R_S(0)$, but this quantity diverges if the potential is
Coulombic at short distances as in the case of QCD.  How does one deal with
this ultraviolet divergence?

4.  {\em How do you take into account higher Fock states?}
While heavy quarkonium is predominantly in the Fock state $\ket{\QQ}$,
the amplitudes for higher Fock states, such as $\ket{\QQg}$
and $\ket{\QQgg}$, are certainly nonzero.  At some level, they
must contribute to annihilation rates.  How can they be taken
into account?

These problems, which blocked progress on the theory of heavy quarkonium
annihilation,
resisted solution for more than a decade.  Surprisingly, the solution to
these problems is rather simple when viewed from the proper perspective.
This perspective is provided by NRQCD, an effective field theory that
makes it relatively easy to separate out effects of different momentum
scales in heavy quarkonium.  Once this framework is developed,
the solutions to all the problems listed above fall into place.

\section{NRQCD and the General Factorization Formula}

\vspace*{-0.7cm}
\subsection{Momentum Scales}
\vspace*{-0.35cm}

Heavy quarkonium is a system that involves several
important momentum scales.  First, there is the heavy quark mass $M$,
which sets the length scale for the $\QQ$ annihilation process.
Then there is the typical momentum $Mv$ of the heavy quark,
which sets the length scale for quarkonium structure.
Next, there is $Mv^2$, the typical kinetic energy of the heavy quark.
This scale is important, because the effects of dynamical gluons
with momenta of order $Mv^2$ can not be taken into account via an effective
potential between the heavy quark and antiquark.  They can however be taken
into account through higher Fock states, such as $\ket{\QQg}$.
Finally, there is $\Lambda_{QCD}$, the scale of nonperturbative effects
associated with light quarks and gluons.  It is the existence of so many
important length scales that makes heavy quarkonium such an
interesting and challenging problem.

If the heavy quark mass $M$ is large enough, than these momentum scales will
be well-separated:
\be
\Lambda_{QCD}^2 \;\ll\; (Mv^2)^2 \;\ll\; (Mv)^2 \;\ll\; M^2 .
\ee
We will refer to $1/M$ as the short-distance scale, and the other 3 scales
will be referred to collectively as long-distance scales.  Our analysis
of heavy quarkonium will be based on separating the effects of the
short-distance scale from those of the long-distance scales.
This separation is very distinct in bottomonium, where the typical value
of $v^2$ is $v^2 \approx 1/10$, and it is reasonably clear in
charmonium, where $v^2 \approx 1/3$.

\vspace*{-0.35cm}
\subsection{NRQCD Lagrangian}
\vspace*{-0.35cm}

A convenient tool for separating the short-distance scale $M$ from the
long-distance scales in heavy quarkonium is nonrelativistic QCD (NRQCD).
This is an effective field theory which was developed by Lepage and
collaborators\cite{caswell-lepage,lmnmh} precisely for this purpose.
It is designed to remove the scale $M$ from the problem, while
precisely reproducing the bound-state dynamics at length scales
of order $1/(Mv)$ or longer.
The lagrangian for NRQCD is
\be
{\cal L}_\NRQCD \;=\; {\cal L}_{\rm light} \;+\; {\cal L}_{\rm heavy} \;+\;
\delta {\cal L} ,
\label{LNRQCD}
\ee
where ${\cal L}_{\rm light}$ is the usual relativistic lagrangian for the
gluons and the light quarks.
The heavy quarks and antiquarks are described by the term
\be
{\cal L}_{\rm heavy}
\;=\; \psidag \, \left( iD_0 + \frac{\Dv^2}{2M} \right)\, \psi
\;+\; \chidag \, \left( iD_0 - \frac{\Dv^2}{2M} \right)\, \chi ,
\label{Lheavy}
\ee
where $\psi$ is a 2-component spinor field that annihilates a heavy quark,
$\chi$ is a 2-component spinor field that creates a heavy antiquark,
and $D_\mu$ is the gauge-covariant derivative.  The lagrangian
${\cal L}_{\rm light} + {\cal L}_{\rm heavy}$ describes ordinary QCD
coupled to a Schr\"odinger field theory for the heavy quarks and antiquarks.
This field theory is nonrenormalizable, and requires an ultraviolet cutoff
$\Lambda$ whose natural scale is of order $M$.  The final term
$\delta{\cal L}$ in the lagrangian (\ref{LNRQCD}) includes all possible local
gauge-invariant counterterms. Their coefficients must be tuned so that
the ultraviolet divergences in NRQCD calculations of
long-distance quantities are cancelled, and so that the residual finite
terms reproduce the corresponding results from full QCD.

The solutions to the heavy-quark field equation
$(i D_0 + \Dv^2/(2 M) ) \psi = 0$
contain a dynamically-generated length scale $Mv$.  This scale is generated
by the nonperturbative balance between the effects of the operators
$D_0$ and $\Dv^2 /2M$, with $D_0 \sim Mv^2$ and $\Dv \sim Mv$.
Thus there is also a dynamically-generated small parameter $v$,
which can be used to analyze the importance of the various terms in the
NRQCD lagrangian\cite{lmnmh}.
By using the field equations, one can show that the
lagrangian ${\cal L}_{\rm light} + {\cal L}_{\rm heavy}$
can be used to calculate matrix elements between quarkonium states up to an
error of order $v^2$.
If higher accuracy is required, one must add the counterterms
\bea
\delta{\cal L}_{\rm bilinear}
&=& \frac{c_1}{8M^3} \psidag (\Dv^2)^2 \psi
\;+\; \frac{c_2}{8M^2} \psidag (\Dv \cdot g \Ev - g \Ev \cdot \Dv) \psi
\nl
&+& \frac{c_3}{8M^2}
	\psidag (i \Dv \times g \Ev - g \Ev \times i \Dv) \cdot \sigmav \psi
\;+\; \frac{c_4}{2M} \psidag (g \Bv \cdot \sigmav) \psi
\nl
&+& {\rm antiquark \; terms} .
\label{Lbilinear}
\eea
The coefficients $c_i$ in (\ref{Lbilinear}) can be determined by demanding
that perturbative scattering amplitudes in NRQCD agree with those in full QCD
to order $v^2$.  Using the lagrangian
${\cal L}_{\rm light} + {\cal L}_{\rm heavy} + \delta {\cal L}$,
matrix elements between heavy quarkonium states can be calculated with
errors of order $v^4$.

\vspace*{-0.35cm}
\subsection{Annihilation in NRQCD}
\vspace*{-0.35cm}

The annihilation of a $\QQ$ pair cannot be described directly
in NRQCD, because annihilation processes such as $Q \Qbar \to g^*$
produce  gluons with small wavelengths of order $1/M$.
However, annihilation can be taken into account indirectly
through its effects on $\QQ$ scattering amplitudes.
At long distances, the amplitude for $Q \Qbar \to g^* \to Q \Qbar$
can be reproduced by a point-like interaction for $\QQ$ scattering.
The effects of annihilation can therefore be taken into account
adding 4-fermion operators to the lagrangian:
\be
\delta {\cal L}_{\rm 4-fermion}
\;=\; \sum_n {f_n(\alpha_s(M)) \over M^{d_n-4}} \; {\cal O}_n ,
\label{L4fermion}
\ee
where the sum is over all possible local 4-fermion operators ${\cal O}_n$
that annihilate and create a heavy-quark-antiquark pair, and $d_n$ is the
scaling dimension of ${\cal O}_n$.  An example of such an operator is
$\psidag \chi \chidag \psi$, which has a scaling dimension of 6.
The coupling constants $f_n(\alpha_s)$ can be computed
by matching perturbative amplitudes for $\QQ$ scattering in NRQCD
with the corresponding amplitudes in full QCD.

According to the optical theorem, the annihilation rate of a $\QQ$ pair
is proportional to the imaginary part of its forward scattering amplitude.
In order to reproduce these imaginary parts in NRQCD,
the coefficients $f_n$ of the 4-fermion operators must have imaginary parts.
These imaginary parts must also reproduce the effects of annihilation
of heavy quarkonium.  Up to a factor of $(-2)$, the decay rate
$\Gamma$ can be identified with the imaginary part of the energy of the state.
Since that energy is the expectation value of the interaction lagrangian,
we can write the annihilation part of the decay rate
of a heavy quarkonium state $H$ as
\be
\Gamma(H) \;=\; 2 \; {\rm Im} \bra{H} {\cal L}_{\rm 4-fermion} \ket{H} .
\ee
Using the expansion (\ref{L4fermion}), this becomes
\be
\Gamma(H) \;=\; \sum_n {2 \; {\rm Im} f_n(\alpha_s(M)) \over M^{d_n-4}}
	\bra{H} {\cal O}_n \ket{H} .
\label{master}
\ee
This is the general factorization formula for the annihilation decay rates of
heavy quarkonium states\cite{bblb}.
It expresses the decay rate as the sum of
long-distance matrix elements $\bra{H} {\cal O}_n \ket{H}$,
multiplied by short-distance coefficients.
The sum is over all possible 4-fermion operators.  All the dynamics
associated with the scale $M$ resides in the short-distance
coefficients, which are proportional to the imaginary parts of
coupling constants in the
NRQCD lagrangian.  All effects from distance scales of order
$1/(Mv)$ or larger are contained in the matrix elements.

\vspace*{-0.35cm}
\subsection{Simplifying the Matrix Elements}
\vspace*{-0.35cm}

As it stands, the general factorization formula is not very useful
because it contains infinitely many terms.  To make it into a
practical tool, we must exploit the dynamically-generated small parameter
$v$ in NRQCD.  As $v \to 0$, each of the matrix elements in (\ref{master})
scales with a definite power of $v$.  At any given order in $v$,
only a finite number of the matrix elements contribute to the decay rate.
The number of independent matrix elements is even smaller, because there
are many relations between the matrix elements that hold at low orders of $v$.
Below, we summarize the methods that can be used to reduce the number
of terms in the general factorization formula.\cite{bblb}

At a given order in $v$, the number of matrix elements can be reduced
to a finite number by using velocity scaling rules for the matrix elements.
These scaling rules consist of scaling rules for the operators and
scaling rules for the probabilities of the Fock states that give the
leading contributions to the matrix elements.  The scaling rules for the
operators are fairly simple,\cite{lmnmh} and can be summarized by
$\Dv \sim v$, $D_0 \sim v^2$, $g \Ev \sim v^3$, and $g \Bv \sim v^4$.
These rules give the scaling behavior of the matrix element
$\bra{H} {\cal O}_n \ket{H}$ if the operator ${\cal O}_n$
annihilates and creates the $\QQ$ pair in the dominant Fock state $\ket{\QQ}$
of the meson $\ket{H}$.  An example is the matrix element
$\bra{\eta_c} \psidag \chi \chidag \psi \ket{\eta_c}$, where the operator
creates and annihilates a $c \bar c$ pair in a color-singlet $^1S_0$ state.
The matrix element $\bra{H} {\cal O}_n \ket{H}$
is suppressed by one extra power of $v^2$
if ${\cal O}_n$ annihilates and creates the $\QQ$ pair in the
$\ket{\QQg}$ Fock state that can be reached by an E1 transition
from the dominant $\ket{\QQ}$ state.  An example is the matrix element
$\bra{h_c} \psidag T^a \chi \chidag T^a \psi \ket{h_c}$, where the operator
creates and annihilates a $c \bar c$ pair in a color-octet $^1S_0$ state.
In the dominant Fock state $\ket{\QQ}$ of the $h_c$, the $\QQ$ pair
is in a color-singlet $^1P_1$ state, but the $h_c$ also has an
amplitude of order $v$ to be in  a $\ket{\QQg}$ Fock state,
with the $\QQ$ pair in a color-singlet $\singS$ state.  Thus the matrix
element is suppressed by a factor of $v^2$ from the
probability of this Fock state.

The matrix elements that contribute to a decay rate to a given order in $v$
are not all independent.
If the operator ${\cal O}_n$ creates and annihilates the $\QQ$ pair in the
dominant Fock state $\ket{\QQ}$, then the matrix element
$\bra{H} {\cal O}_n \ket{H}$ can be simplified by using
the vacuum-saturation approximation.  For example,
\be
\bra{\eta_c} \psidag \chi \chidag \psi \ket{\eta_c} \;=\;
\bra{\eta_c} \psidag \chi \ket{0} \bra{0} \chidag \psi \ket{\eta_c}
	\Bigg( 1 \;+\; O(v^4) \Bigg) .
\ee
The vacuum-saturation approximation provides relations between the
matrix elements that contribute to decays into light hadrons and those
that contribute to decays into electromagnetic final states.
In NRQCD, this is a controlled approximation with a relative error
that scales like $v^4$.
There are also relations between matrix elements that follow from
heavy-quark spin symmetry.  This is an exact symmetry of the
lagrangian ${\cal L}_{\rm light} + {\cal L}_{\rm heavy}$.  Thus the
relations implied by heavy-quark spin symmetry are correct up to
errors of relative order $v^2$.  An example of a relation that follows from
heavy-quark spin symmetry is
\be
\bra{0} \chidag \psi \ket{\eta_c} \;=\;
\epsilonv \cdot \bra{0} \chidag \psi \ket{\psi(\epsilonv)}
		\Bigg( 1 \;+\; O(v^2) \Bigg) .
\ee
where $\epsilonv$ is the polarization vector of the $\psi$.

\section{Applications}

\vspace*{-0.7cm}
\subsection{S-waves at Leading Order in $v^2$}
\vspace*{-0.35cm}

In previous calculations for S-wave states, the annihilation decay rates
were assumed to factor into short-distance coefficients multiplied by
$|R_S(0)|^2$, where $R_S(0)$ is  the radial wavefunction at the origin.
This long-distance factor was assumed to be the same for decays of the
$0^{-+}$ state and the $1^{--}$ state.  It was also assumed to be the
same for decay rates into electromagnetic final states and for inclusive
decay rates into hadrons.  Our general factorization theorem verifies that
these assumptions are correct at leading order in $v^2$.
At leading order in $v^2$, the decay rates of the
$0^{-+}$ state $\eta_c$ and the $1^{--}$ state $\eta_c$ can be written
\bea
\Gamma(\eta_c \to \LH)
&=& {F_1(\singS) \over M^2} \; \Big| \RS \Big|^2 ,
\label{Getalh0}
\\
\Gamma(\psi \to \LH)
&=& {F_1(\tripS) \over M^2} \; \Big| \RS \Big|^2 ,
\label{Gpsilh0}
\\
\Gamma(\eta_c \to \gamma \gamma)
&=& {F_{\gam}(\singS) \over M^2} \; \Big| \RS \Big|^2 ,
\label{Getagg0}
\\
\Gamma(\psi \to e^+ e^-)
&=& {F_{\lep}(\tripS) \over M^2} \; \Big| \RS \Big|^2 .
\label{Gpsiee0}
\eea
In (\ref{Getalh0}) and (\ref{Gpsilh0}), the vacuum-saturation
approximation has been used to replace the matrix elements
$\bra{\eta_c} \psidag \chi \chidag \psi \ket{\eta_c}$ and
$\bra{\psi} \psidag \sigmav \chi \cdot \chidag \sigmav \psi \ket{\psi}$
by the same factor $|\RS|^2$ that appears in the electromagnetic
decay rates (\ref{Getagg0})
and (\ref{Gpsiee0}).  The radial wavefunctions at the origin
for the $\psi$ and $\eta_c$ are given by the matrix elements
\bea
\Reta &\equiv&
\sqrt{{2 \pi \over 3}} \; \bra{0} \chidag \psi \ket{\eta_c} ,
\label{Reta}
\\
\Rpsi &\equiv& \sqrt{{2 \pi \over 3}} \; \epsilonv \cdot
\bra{0} \chidag \sigmav \psi \ket{\psi(\epsilonv)} \;.
\label{Rpsi}
\eea
At leading order in $v^2$, we can use heavy-quark spin symmetry to
equate $\RS$ in (\ref{Getalh0})--(\ref{Gpsiee0}) with $\Reta$, or with
$\Rpsi$, or with their weighted average:  $\RS = (\Reta + 3 \Rpsi)/4$.
At this level of accuracy, the only advantage of the NRQCD formalism
is that it provides a rigorous nonperturbative definition of the
long-distance factor $|\RS|^2$, so that it can be calculated with
lattice simulations.  It has been calculated\cite{tl,bks}
to leading order in $v^2$.
The short-distance coefficients $F_i$ in the factorization formulas
(\ref{Getalh0})--(\ref{Gpsiee0}) are proportional to the imaginary
parts of coupling constants in the NRQCD lagrangian.
They have all been calculated to next-to-leading order in $\alpha_s$.

\vspace*{-0.35cm}
\subsection{S-waves at Next-to-leading Order in $v^2$}
\vspace*{-0.35cm}

At next-to-leading order in $v^2$, the factorization formula for S-wave
states becomes less trivial, with 3 independent long-distance factors:
\bea
\Gamma(\eta_c \to \LH)
&=& {F_1(\singS) \over M^2 } \; \Big| \Reta \Big|^2
\;+\; {G_1(\singS) \over M^4} \; {\rm Re}(\RS{}^*\, \ddRS) ,
\label{Getalh2}
\\
\Gamma(\psi \to \LH)
&=& {F_1(\tripS) \over M^2} \; \Big| \Rpsi \Big|^2
\;+\; {G_1(\tripS) \over M^4} \; {\rm Re}(\RS{}^*\, \ddRS) ,
\label{Gpsilh2}
\\
\Gamma(\eta_c \to \gamma \gamma)
&=& {F_{\gam}(\singS) \over M^2} \; \Big| \Reta \Big|^2
\;+\; {G_{\gam}(\singS) \over M^4} \; {\rm Re}(\RS{}^*\, \ddRS) ,
\label{Getagg2}
\\
\Gamma(\psi \to e^+ e^-)
&=& {F_{\lep}(\tripS) \over M^2} \; \Big| \Rpsi \Big|^2
\;+\; {G_{\lep}(\tripS) \over M^4} \; {\rm Re}(\RS{}^*\, \ddRS) .
\label{Gpsiee2}
\eea
In the first terms on the right sides of (\ref{Getalh2})--(\ref{Gpsiee2}),
the radial wavefunctions at the origin $\Reta$ and $\Rpsi$
can not be equated with their weighted average $\RS$ without loss of accuracy.
The third independent matrix element $\ddRS$
can be interpreted as a renormalized laplacian of the wavefunction
at the origin:
\be
\ddRS \;\equiv\; \sqrt{2 \pi \over 3} \;
\bra{0} \chidag \gradv^2 \psi \ket{\eta_c} \Bigg|_{\rm Coulomb}.
\label{R2etaC}
\ee
The short-distance coefficients $F_i$ in (\ref{Getalh2})--(\ref{Gpsiee2})
are the same as in (\ref{Getalh0})--(\ref{Gpsiee0}), and are known to
next-to-leading order in $\alpha_s$.  The short-distance coefficients
$G_i$ can be extracted at leading order in $\alpha_s$ from
calculations by Keung and Muzinich\cite{km}, who developed a
phenomenological model for taking into account relativistic corrections to
quarkonium decays.  A phenomenological analysis of S-wave decays,
including the relativistic corrections, is in progress\cite{bblz}.

\vspace*{-0.35cm}
\subsection{P-waves at Leading Order in $v^2$}
\vspace*{-0.35cm}

The correct factorization formulas for decays of P-wave states\cite{bbla}
are nontrivial even at leading order in $v^2$:
\bea
\Gamma(h_c \to \LH)
&=& {F_1(\singP) \over M^4} \; \Big| \dRP \Big|^2
\;+\; {F_8(\singS) \over M^2} \; \langle {\cal O}_8 \rangle ,
\label{hchad}
\\
\Gamma(\chic{J} \to \LH)
&=& {F_1(\tripP{J}) \over M^4} \; \Big| \dRP \Big|^2
\;+\; {F_8(\tripS) \over M^2} \; \langle {\cal O}_8 \rangle ,
\quad \; J=0,1,2,
\label{chihad}
\\
\Gamma(\chic{J} \to \gamma \gamma)
&=& {F_{\gam}(\tripP{J}) \over M^4} \;	\Big| \dRP \Big|^2 ,
	\quad \; J = 0,2.
\label{chigg}
\eea
There are two independent long-distance
factors that appear in these decay rates.
The long-distance factor $|\dRP|^2$ can be identified with
the square of the derivative of the wavefunction at the origin
for the $h_c$ or any of the other P-states:
\be
\dRP \;\equiv\; \sqrt{2 \pi \over 9} \; \epsilonv \cdot
\bra{0} \chidag \gradv \psi \ket{h_c(\epsilonv)} \Bigg|_{\rm Coulomb}.
\label{RhC}
\ee
The second long-distance factor in (\ref{hchad}) and (\ref{chihad}) is the
matrix element of a 4-fermion operator in NRQCD:
\be
\langle {\cal O}_8 \rangle
	\;\equiv\; \bra{h_c} \psidag T^a \chi \chidag T^a \psi \ket{h_c} .
\label{O8}
\ee
The operator in the matrix element annihilates and creates a $\QQ$
pair in a color-octet S-wave state, and thus measures the probability
density for the $\QQ$ pair to be in a color-octet S-wave state
at the origin.  This term in the factorization formula represents
a contribution to the decay rate from the $\ket{\QQg}$ component
of the quarkonium wavefunction.  The matrix element
$\langle {\cal O}_8 \rangle$ depends logarithmically on the ultraviolet
cutoff $\Lambda$ of NRQCD, which can be interpreted as an arbitrary
factorization scale in the full theory:
\be
\Lambda {d \ \over d \Lambda} \langle {\cal O}_8 \rangle
	\;=\; {8 \alpha_s \over 3 \pi^2 M^2} | \dRP |^2 .
\label{rge}
\ee
The $\Lambda$-dependence of $\langle {\cal O}_8 \rangle$ must be cancelled
by the coefficients $F_i$ in the factorization formulas (\ref{hchad})
and (\ref{chihad}).  The logarithmic dependence of these coefficients on
$\Lambda$ explains the infrared divergences discovered by
Barbieri {\it et al.} in their pioneering calculations of P-wave
decays\cite{barbc,barbe}.

The short-distance coefficients $F_1(\tripP{0})$, $F_1(\tripP{2})$,
$F_8(\singS)$, and $F_8(\tripS)$ in (\ref{hchad})--(\ref{chigg})
are known\cite{bbla} to leading order in $\alpha_s$,
while $F_{\gam}(\tripP{0})$ and  $F_{\gam}(\tripP{2})$ have been calculated
to next-to-leading order\cite{barbe}.  Accurate lattice calculations
of the long-distance factors are not yet available.  Fortunately the
long-distance factors can be determined phenomenologically.  Experiment E760
at Fermilab\cite{E760}, which studies the resonant production of
charmonium in $p {\bar p}$ collisions, has provided precise measurements
of the total widths of the $\chi_{c1}$ and $\chi_{c2}$\cite{E760had}.
They can be used
to determine $\dRP$ and $\langle {\cal O}_8 \rangle$, and then the remaining
four P-wave decay rates can be predicted.  The prediction for
$\chi_{c2} \to \gamma \gamma$ is consistent with recent measurements
by E760\cite{E760rad} and by CLEO\cite{CLEO}.
The predictions for the total widths of the $h_c$ and $\chi_{c0}$
and for the rate for $\chi_{c0} \to \gamma \gamma$ should be tested in
the followup experiment E835.

\section{Summary}

A rigorous theory for the annihilation decay rates of heavy quarkonium
has recently been developed.  The inclusive decay rate of a quarkonium
state $H$ into light hadrons satisfies a general factorization formula:
\begin{equation}
\Gamma(H) \;=\; \sum_n {F_n(\alpha_s(M)) \over M^{d_n - 4}}
\langle H | {\cal O}_n | H \rangle .
\end{equation}
The sum is over all 4-fermion operators in NRQCD that create and annihilate
both a quark and an antiquark, but at any given order in $v^2$, there are
only a finite number of operators that contribute.  The long-distance factors
$\langle H |{\cal O}_n | H \rangle$ are well-defined
matrix elements that can be calculated numerically using lattice simulations
of NRQCD.  The short-distance coefficients $F_n$ can be calculated as
perturbation expansions in $\alpha_s(M)$.  Using this general factorization
theorem, it is now possible to calculate the annihilation decay rates
of heavy quarkonium from first principles, with the only inputs being the
heavy-quark mass $M$ and the QCD coupling constant.

\section{Acknowledgements}
This work was supported in part by the U.S. Department of Energy,
Division of High Energy Physics, under Grant DE-FG02-91-ER40684.
I wish to thank my collaborators G.T. Bodwin and G.P. Lepage
for sharing their insights into the physics of heavy quarkonium.

\end{document}